# Heat conduction in molecular transport junctions

Michael Galperin and Mark A. Ratner
Department of Chemistry, Northwestern University, Evanston, 60208
and
Abraham Nitzan
School of Chemistry, Tel Aviv University, Tel Aviv, 69978, Israel

## Abstract

Heating and heat conduction in molecular junctions are considered within a general NEGF formalism. We obtain a unified description of heating in current carrying molecular junctions as well as the electron and phonon contributions to the thermal flux, including their mutual influence. Ways to calculate these contributions, their relative importance and ambiguities in their definitions are discussed. A general expression for the phonon thermal flux is derived and used in a new "measuring technique", to define and quantify 'local temperature' in nonequilibrium systems. Superiority of this measuring technique over the usual approach that defines effective temperature using the equilibrium phonon distribution is demonstrated. Simple bridge models are used to illustrate the general approach, with numerical examples.

# 1. Introduction

Research in molecular electronics, in particular the study of electron transport through a molecular system coupled to metal and/or semiconductor leads, is motivated both by scientific challemges and by its potential for complementing existing Si based electronics by new molecular size devices. An intriguing issue in this field is the interplay between electron transport and nuclear motions in the non-equilibrium junction. Inelastic effects in the current/voltage response are directly observed in inelastic electron tunneling spectroscopy (IETS) experiments and provide an invaluable diagnostic tool for junction composition and structure, most importantly for confirming the presence of molecule(s) in the junction. They also manifest themselves in current-induced chemistry and/or molecular motion in single-molecule STM junctions. Vibrational features were reported both in the Coulomb blockade and Kondo regimes of junction transport. The field continues to be very active in both experimental and theoretical studies.

An important consequence of electron-vibration interaction in junction transport is heat generation, i.e. energy transfer to the underlying nuclear motions. In balance with the process of heat dissipation – conduction of thermal energy away from the junction - this has important implications on the issue of junction stability. These processes have attracted much experimental[1-6] and theoretical.[7] [8-15] [16] attention.

The problem of heat generation in a current carrying junction is concerned with the fraction of available power, $I\Phi$ (for a junction carrying current $I$ under a potential bias $\Phi$), that is left as heat in the junction.[7-9, 17-20] In junctions where the molecular wire is suspended between the source and drain leads, when the heating issue is most acute, heat dissipation is directly related to the process of heat conduction through the wire. In another guise, heat transport through a chain of coupled oscillators connecting two thermal baths, this problem has attracted much attention on its own,[6, 13, 14, 21-35] focusing on issues of dimensionality, the validity of the heat equation in transport through harmonic chains and the effects of the quantum nature of the conducting modes. The latter issue is of particular interest because quantum mechanical considerations imply a limited number of vibrational modes available for heat conduction and the existence of a quantum of heat conduction. Applications to molecular wires[10-16] address this issue independently from the problem of heat generation. A unified semiclassical approach is provided by using phenomenological kinetic equations.[36-43] Indeed, the two issues should be addressed together for several reasons. First, in contrast to heat transport in coupled oscillator chains connecting two boson baths, heating in current carrying molecular wires is a transport phenomenon involving electrons, phonons and their mutual coupling. Even the definition of heating should be addressed carefully, with the need to distinguish between energy transfer from the electronic to the nuclear subsystems and energy randomization within the electronic subsystem due to electron-electron interaction.† Second, heat dissipation in such systems can result from coupling not only to external phonon baths but to the thermal manifolds of metallic electron states. Third, a system with coupled electron and phonon transport processes is characterized also by cross-correlations, manifested in thermoelectric transport phenomena. Finally, for a molecule connecting two metallic leads at different temperatures, it is of interest to ask whether the contributions of the electron and phonon subsystems to the overall heat conduction are separable, and to examine their relative magnitudes.

Following Landauer, most of theoretical work on nanojunction transport is done within a scattering theory approach, which disregards the contacts and their influence on the scattering channels as well as the mutual influences of the electron and phonon subsystems on each other. This approach is known to fail in particular cases.[44, 45] A more general non-equilibrium Green function (NEGF) approach for thermal transport, which takes care of the aforementioned issues, was pioneered in work by Datta and coworkers.[17-19] In particular, as shown below, it can deal with both electron and phonon assisted thermal transport in a unified way. The importance of such a self consistent approach is expected to be especially pronounced in the strongly non-equilibrium junction (large source-drain voltage) situation, when the electron flux has enough energy to strongly excite vibrational modes, and in the case of strong electron-phonon coupling that is often encountered in resonant tunneling situations. This paper introduces and develops such a NEGF based approach to describe thermal transport through molecular junctions and applies it to the issues mentioned above using simple model calculations. Section 2 introduces the model used to simulate the metal-molecule-metal junction and its relevant interaction parameters and Section 3 describes the procedure used to calculate electron and phonon assisted‡ thermal transport. In section 4 we discuss the relevance and meaning of the 'junction temperature' under non-equilibrium operating conditions. Results of model calculations are presented in Section 5. Section 6 concludes.

## 2. Model

We consider a two-terminal junction made of two leads (L for left and R for right), represented by free electron reservoirs each in its own thermal equilibrium, interconnected by a bridging molecular system. We will implicitly assume the existence of a third lead, a gate, capacitively coupled to the junction that can be used to change the energy of molecular levels relative to the

---

† This process is usually disregarded in molecular wires where only a few electronic states (if any) are delocalized and the spacing between electronic levels is often large relative to wire temperature, but can mark the difference between small molecular wires and, say, metallic carbon nanotubes.
‡ We use the term phonon quite generally, to name a quantized vibrational mode.

Fermi energy. In what follows, we refer to the molecular bridge (possibly with a few atoms on both sides constituting together an extended molecule) as our system. Nuclear motions (of both the molecule and solvent) are described as harmonic normal modes and are divided into two groups. The primary group includes local vibrations that are driven out of equilibrium in the course of the transport process. The secondary phonons represent the environment, taken to be coupled linearly to the primary group and assumed to be in thermal equilibrium. The assignment of phonons to these groups depends on the problem. For the study of heat generation in a current carrying junction we associate with the primary group phonons that directly interact with the electronic states of the bridge. They are driven by the non-equilibrium electronic system while concurrently relaxing by their coupling to the thermal environment. When discussing heat transport through a molecular bridge connecting two thermal baths, all the bridge phonons are parts of our system, i.e. in the primary group, irrespective of their coupling to the electronic subsystem. The secondary groups include phonons in the two thermal reservoirs. In the present discussion these two criteria are assumed equivalent, i.e. it is assumed that the bridge phonons are those that couple to the electronic process.

The model Hamiltonian is divided into zero-order and interaction part

$$\hat{H} = \hat{H}_0 + \hat{V} \qquad (1)$$

where $\hat{H}_0$ represents the non-interacting subsystems. In the second quantization it reads (here and below we use $\hbar = 1$ and $e = 1$)

$$\hat{H}_0 = \sum_{k \in L, R} \varepsilon_k \hat{c}_k^\dagger \hat{c}_k + \sum_{i,j=1}^{N} H_{ij}^M \hat{d}_i^\dagger \hat{d}_j + \sum_{i=1}^{N} V_i^{ext} \hat{d}_i^\dagger \hat{d}_i + \sum_\alpha \omega_\alpha \hat{a}_\alpha^\dagger \hat{a}_\alpha + \sum_\beta \omega_\beta \hat{b}_\beta^\dagger \hat{b}_\beta \qquad (2)$$

$\hat{b}_\beta \left( \hat{b}_\beta^\dagger \right)$ and $\hat{a}_\alpha \left( \hat{a}_\alpha^\dagger \right)$ are annihilation (creation) operators for secondary and primary vibrational normal modes respectively. $\hat{c}_k \left( \hat{c}_k^\dagger \right)$ is annihilation (creation) of electron in the leads, and $\hat{d}_j \left( \hat{d}_j^\dagger \right)$ is corresponding operator(s) for electron states on the bridge. The five terms on the right of Eq. (2) represent electrons in the left and right leads, electrons on the molecule (in a representation defined by a basis of $N$ single electron orbitals), an external potential, primary vibrations and secondary vibrations, respectively. The lead electrons and the secondary phonons are assumed to be in their own equilibrium, defined by the temperature and the electron chemical potential of each lead. The molecular system is driven out of equilibrium when the different baths are not in equilibrium with each other. The single electron basis chosen to represent the molecular electronic system can correspond to atomic or molecular orbitals, lattice points, plane waves, or any other convenient basis. The external potential term can represent a gate potential $V_i^{ext} = V_g$. Below we will often use a single-level molecular model that corresponds to the molecular orbital of energy $\varepsilon_0$ relevant to the energy range of interest. We will also consider a single primary vibrational mode of frequency $\omega_0$, so that $\sum_m \omega_m \hat{a}_m^\dagger \hat{a}_m \to \omega_0 \hat{a}_0^\dagger \hat{a}_0$ and $\sum_{i,j} H_{ij}^M \hat{d}_i^\dagger \hat{d}_j + \sum_i V_i^{ext} \hat{d}_i^\dagger \hat{d}_i \to \left(\varepsilon_0 + V_0^{ext}\right) \hat{d}_0^\dagger \hat{d}_0$. To simplify notation we will disregard $V_0^{ext}$ and assume that $\varepsilon_0$ can be varied.

The interacting part of the Hamiltonian couples between the above subsystems.

$$\hat{V} = \sum_{k \in L, R, i} \left( V_{ki} \hat{c}_k^\dagger \hat{d}_i + V_{ik} \hat{d}_i^\dagger \hat{c}_k \right) + \sum_{i,j;\, \alpha} M_{ij}^\alpha \hat{Q}_\alpha^a \hat{d}_i^\dagger \hat{d}_j + \sum_{\alpha,\beta} U_{\alpha\beta} \hat{Q}_\alpha^a \hat{Q}_\beta^b \qquad (3)$$

Here $\hat{Q}_\alpha^a$ and $\hat{Q}_\beta^b$ are vibration displacement operators

$$\hat{Q}_\alpha^a = \hat{a}_\alpha + \hat{a}_\alpha^\dagger \qquad \hat{Q}_\beta^b = \hat{b}_\beta + \hat{b}_\beta^\dagger \qquad (4)$$

For future reference we also introduce the corresponding momentum operators

$$\hat{P}_\alpha^a = -i\left(\hat{a}_\alpha - \hat{a}_\alpha^\dagger\right) \qquad \hat{P}_\beta^b = -i\left(\hat{b}_\beta - \hat{b}_\beta^\dagger\right) \qquad (5)$$

In Eq. (3), the first term on the right couples between the free electron reservoirs in the leads and the molecular electronic subsystem. The second term is the vibronic coupling between electrons on the molecule and the primary vibrations, which is taken to be linear in the vibrational displacements. The third term represents bilinear coupling between primary vibrational modes and the secondary phonons. Such bilinear coupling is appropriate in a representation where the primary modes correspond to vibrations localized on the molecular bridge.

The physics of the model is dominated by several characteristic parameters: $\Delta E$ – the spacing between the leads Fermi energies and the energy $\varepsilon_0$ of the closest molecular orbital (HOMO and/or LUMO); $\Gamma$ – the broadening of the molecular level due to electron transfer interaction with the leads; $M$ – the electron-primary vibration coupling and $\omega_0$ – the vibrational frequency of the primary mode. In addition, the bias potential $\Phi$ determines the possibility to pump energy into vibrational modes by the threshold condition $e\Phi \geq \hbar\omega_0$. We expect that the effects of electron-phonon interaction on the junctions dynamics will be considerable above this threshold, in particular when the timescale for the electron dynamics (that can be estimated as $\hbar |\Delta E + i\Gamma|^{-1}$) is comparable to that of the primary vibrations, $\omega_0^{-1}$, and when the vibronic coupling $|M|$ is not too small relative to the molecule-lead coupling as determined by $\Gamma$. Below we will also distinguish between resonant and non-resonant electron transmission, for which different levels of mathematical treatment are needed.

## 3. Method

The mathematical objects of interest in the NEGF approach to electron and phonon transport are the corresponding Green functions (GFs) for the electron and the primary phonons

$$G_{ij}(\tau,\tau') = -i\langle T_c \hat{d}_i(\tau)\hat{d}_j^\dagger(\tau')\rangle \qquad (6)$$

$$D_{\alpha\alpha'}(\tau,\tau') = -i\langle T_c \hat{Q}_\alpha^a(\tau)\hat{Q}_{\alpha'}^{a,\dagger}(\tau')\rangle \qquad (7)$$

where $T_c$ is the contour-time ordering operator (later times on the left) on the Keldysh contour. Approximate ways to calculate these GFs were described in our previous publications for the cases of weak[46] and strong[47] electron-phonon interaction. Once the GFs have been calculated they can be used for calculation of thermal transport as described below.

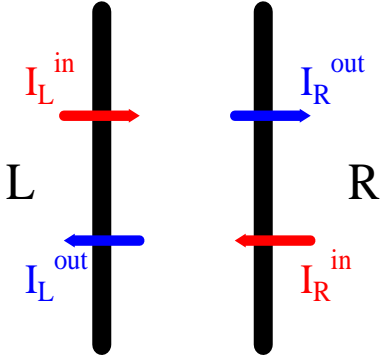

**Figure 1 Electron fluxes through the junction interfaces. L and R represent the left and right leads, respectively.**

Within the NEGF approach one can distinguish between the incoming and outgoing electron fluxes at each molecule-lead (L and R) interface (see Figure 1). The net flux into the molecule at contact $K$ ($K = L,R$) is

$$I_K = I_K^{in} - I_K^{out} \qquad (8a)$$

$$I_K^{in,out} = \frac{1}{\hbar}\int_{-\infty}^{+\infty}\frac{dE}{2\pi}i_K^{in,out}(E) \qquad (8b)$$

where

$$\begin{aligned}i_K^{in}(E) &= \text{Tr}\left[\Sigma_K^<(E)G^>(E)\right]\\ i_K^{out}(E) &= \text{Tr}\left[\Sigma_K^>(E)G^<(E)\right]\end{aligned} \qquad (9)$$

Here $\Sigma_K^{<,>}$ are the lesser/greater projections of the self-energy due to coupling to the contact $K$

$$\Sigma_K^<(E) = if_K(E)\Gamma_K(E)$$
$$\Sigma_K^>(E) = -i[1-f_K(E)]\Gamma_K(E) \qquad (10)$$

with $f_K(E) = \left[\exp\left((E-\mu_K)/k_B T_K\right)+1\right]^{-1}$ is the Fermi distribution in the contact $K$ and

$$[\Gamma_K]_{ij}(E) = 2\pi \sum_{k \in K} V_{ik}V_{kj}\delta(E-\varepsilon_k) \qquad (11)$$

Following Lake and Datta[18, 19] we also introduce the corresponding energy current at each interface

$$\left(J_E^{el}\right)_{L,R}^{in,out} = \int_{-\infty}^{+\infty} \frac{dE}{2\pi} E\, i_{L,R}^{in,out}(E) \qquad (12)$$

and the net energy flux into the junction at the interface $K$

$$J_{E,K}^{el} = \int_{-\infty}^{+\infty} \frac{dE}{2\pi} E\, i_K(E); \qquad i_K(E) = i_K^{in}(E) - i_K^{out}(E) \qquad (13)$$

The net rate of energy change in the bridge electronic subsystem can be expressed in terms of these fluxes in the form,

$$J_{\Delta E}^{el} = J_{E,L}^{el} + J_{E,R}^{el} \qquad (14)$$

(The subscript $\Delta E$ is used to emphasize that that this is the rate of *electronic energy change* on the bridge, not an energy flux through the bridge). This net flux is zero in the absence of additional routes for energy dissipation (e.g. phonons) on the bridge. In contrast, the electronic heat flux out of the lead $K = L, R$, given by[48]

$$J_{Q,K}^{el} = \int_{-\infty}^{+\infty} \frac{dE}{2\pi}(E-\mu_K)\, i_K(E) \qquad (15)$$

does not represent a conserved quantity. Consequently away from equilibrium when $\mu_L \neq \mu_R$ the difference $J_{Q,L}^{el} - J_{Q,R}^{el}$ does not vanish even in the absence of system phonons.[§] It represents the net generation rate of Joule heat in the current carrying system.

     In the presence of bridge vibrations one has to take into account energy and heat transport also via the phonon subsystem, as well as the effect of electron-phonon coupling. Two issues are of particular interest, one pertaining to heat generation on the junction and the other to heat transport through the junction:

(a) Electron-phonon interaction causes energy exchange between the electron and the (primary) phonon subsystems on the bridge. In a biased junction the applied voltage is the energy source, and we can regard the net energy flux from electrons to primary phonons as the rate of heat generation on the bridge. Energy conservation implies that this rate is given by $J_{\Delta E}^{el}$ of Eq. 14. We conclude that the heat generation rate on the molecular bridge is given by

$$J_Q = J_{\Delta E}^{el} \qquad (16)$$

---

[§] One still assumes the existence of a dissipation mechanism that keeps the leads in their corresponding equilibrium states.

(b) Phonons obviously contribute to heat transport through the bridge, as do electrons in metal-molecule-metals contacts. Electron-phonon interaction can play an important role in this heat transport process, as indicated by the fact that the only applicable conservation law is that of the total (electronic and nuclear) energy.

Note that for transport by phonons energy and heat fluxes are equivalent, because in the absence of particle conservation there is no chemical potential for phonons. A general quantum expression for phonon thermal flux within NEGF can be derived in complete analogy with original derivation for electron current;[49, 50] the only difference being the carrier statistics. The phonon thermal flux at the interface $K = L, R$ is obtained in the form (see Appendix A and Refs.[51, 52])

$$J_K^{ph} = -\int_0^\infty \frac{d\omega}{2\pi} \omega \, \mathrm{Tr}\left[\Pi_K^{ph,<}(\omega) D^>(\omega) - \Pi_K^{ph,>}(\omega) D^<(\omega)\right] \qquad (17)$$

where Tr stands for summing over all primary phonons, $\Pi_K^{ph,>}$ and $\Pi_K^{ph,<}$ are the greater and lesser self energy (SE) matrices of the primary vibrations due to their coupling to the thermal bath of the contact $K$

$$\begin{aligned}
\left[\Pi_K^{ph,<}(\omega)\right]_{\alpha\alpha'} &= -i\,\Omega_{\alpha\alpha'}^K(\omega)\,F_K(\omega) \\
\left[\Pi_K^{ph,>}(\omega)\right]_{\alpha\alpha'} &= -i\,\Omega_{\alpha\alpha'}^K(\omega)\,F_K(-\omega)
\end{aligned} \qquad (18)$$

with

$$\begin{aligned}
F_K(\omega) &= \begin{cases} N_K(|\omega|) & \omega > 0 \\ 1 + N_K(|\omega|) & \omega < 0 \end{cases} \\
\Omega_{\alpha\alpha'}^K(\omega) &= 2\pi \sum_{\beta \in K} U_{\alpha\beta} U_{\beta\alpha'} \delta(\omega - \omega_\beta)
\end{aligned} \qquad (19)$$

and where

$$N_K(\omega) = N_{eq}(\omega, T_K) \equiv \left[\exp(\omega/k_B T_K) - 1\right]^{-1} \qquad (20)$$

is the Bose-Einstein distribution in the contact $K$.

The bilinear form (last term in Eq. (3)) of the coupling between system (primary) and bath (secondary) phonons is convenient in that it leads to expression (17) which is exact within the non-crossing approximation for any order of the bridge-bath interaction. It holds in principle also in the presence of electron-electron and nuclear anharmonic interactions on the bridge. This form is however not very realistic for the molecule-metal contact, where the Debye frequency in the solid is often smaller than characteristic molecular vibrational frequencies. In this case creation or annihilation of a phonon in the bridge is accompanied by a multiphonon process in the thermal bath. A possible workaround is to introduce an effective exponential density of the thermal bath modes which, coupled bilinearly to the molecule, mimic the effect of the multiphonon process. We do it by using the following model for $\Omega^K(\omega)$:

$$\Omega^K(\omega) = \Omega_0^K \left[\frac{\omega}{\omega_c^K}\right]^2 \exp\left\{2\left(1-\frac{\omega}{\omega_c^K}\right)\right\} \tag{21}$$

where $\Omega_0^K$ and $\omega_c^K$ are constants. In particular $\omega_c^K$ is the cutoff frequency for the reservoir K. The results given and discussed below were obtained using this model. Alternatively one can consider more realistic molecule-bath interactions, e.g.,

$$\sum_\alpha U_\alpha \, F\left(\{\hat{Q}_\beta^b\}\right) \hat{Q}_\alpha^a; \quad F\left(\{\hat{Q}_\beta^b\}\right) \equiv \prod_\beta F_\beta\left(\hat{Q}_\beta^b\right) \tag{22}$$

where $F_\beta\left(\hat{Q}_\beta^b\right)$ is a physically motivated function of the thermal phonon coordinates[53] that reflects the short range nature of inter-nuclear interaction. Such model can not be solved exactly, however, and a possible way to handle it is discussed in Appendix B.

Eq. (17) is exact within the non-crossing approximation, limited only by the requirement that the reservoirs can be represented as collections of independent harmonic modes coupled bilinearly to the molecular bridge. In particular it can be used for the thermal phonon current at the interface between the molecular bridge and the contact K in the presence of anharmonic as well as electron-phonon interactions on the bridge. Consider first a purely harmonic bridge with no electron-phonon interactions. In this case the Keldysh equation for the lesser and greater phonon GFs reads

$$D^{>,<}(\omega) = D^r(\omega)\left[\Pi_L^{>,<}(\omega) + \Pi_R^{>,<}(\omega)\right] D^a(\omega) \tag{23}$$

Using this and Eqs. (18)-(20) in (17) leads to

$$J^{ph} = \int_0^\infty \frac{d\omega}{2\pi} \omega \, Tr\left[\Omega^L(\omega) D^r(\omega) \Omega^R(\omega) D^a(\omega)\right] \left(N_L(\omega) - N_R(\omega)\right) \tag{24}$$

This is a Landauer-type expression that was obtained and studied previously for the phonon mediated heat current in a harmonic junction, without electrons and electron-phonon coupling, by several workers.[12, 28, 54].

Next consider the effect of electron-phonon interaction on the phonon heat current, starting again from Eq. (17) and including the effect of this interaction in the calculation of the GFs. In general no further simplification is possible, however a simple result can be derived in the special (usually unrealistic) case where electron-phonon interaction is present and, e.g., may cause decoherence within the electron and phonon sub-spaces, however no energy exchange between these subsystems takes place on the bridge. In this case at steady-state the phonon thermal flux is the same throughout the junction including the $L$ and $R$ interfaces. This can be used to symmetrize Eq. (17) following a procedure similar to that used by Meir and Wingreen in the electronic case,[49] (using the additional assumption $\Omega^L(\omega) = x \cdot \Omega^R(\omega)$ for any $\omega$; see Appendix C) which leads to

$$J^{ph} = \frac{1}{\hbar} \int_0^\infty \frac{d\omega}{2\pi} \omega \, Tr\left[\Omega^L(\omega) D^r(\omega) \Omega^R(\omega) D^a(\omega)\right] \left(N_L(\omega) - N_R(\omega)\right)$$
$$+ \frac{1}{\hbar} \int_0^\infty \frac{d\omega}{2\pi} \omega \, Tr\left[\frac{\Omega^L(\omega) \Omega^R(\omega)}{\Omega(\omega)} D^r(\omega) \Omega_{el}(\omega) D^a(\omega)\right] \left(N_L(\omega) - N_R(\omega)\right) \quad (25)$$

where $\Omega = \Omega_L + \Omega_R$ and where $\Omega_{el}(\omega) = -2\,\text{Im}\left[\Pi^r_{el}(\omega)\right]$ is the imaginary part of the retarded projection of the contribution to the primary phonons SE associated with their coupling to the electronic subsystem. This contribution to the primary phonons SE was argued to be dominant in the lifetime broadening of these phonons as observed by inelastic electron tunneling spectroscopy at low temperatures.[55]

The result (25) extends Eq. (24) with additive contribution(s) associated with these interaction(s). Note that the same formal form, Eq. (25), is obtained in the more general case that includes anharmonic interactions between bridge phonons, except that $\Omega_{el}(\omega)$ is replaced by a more general function that includes also the effects of such interactions. We will not discuss this issue further in this paper.

In general, energy exchange between electrons and phonons on the bridge cannot be disregarded, and Eq. (17) has to be used directly, treating the mutual influence of these bridge subsystems in a self-consistent manner. Procedures for such self consistent calculations that were developed by us before,[46, 47] yield the corresponding electron and phonon GFs and SEs. They can be used to evaluate the electronic current and the electron and phonon energy/heat fluxes as well as cross correlation effects. Results of such calculations are reported in Section 5.

As stated above, most of the calculations described below are done for the simplest model of single bridge level $\varepsilon_0$ coupled to one vibrational mode $\omega_0$ with leads taken each at its own equilibrium characterized by electrochemical potentials $\mu_L$ and $\mu_R$ and temperatures $T_L$ and $T_R$. As in Ref. [46] we assume that the source-drain voltage $\Phi$ across the junction shifts the electrochemical potentials in the leads relative to $\varepsilon_0$ according to

$$\mu_L = E_F + \frac{\Gamma_R}{\Gamma} e\Phi, \quad \mu_R = E_F - \frac{\Gamma_L}{\Gamma} e\Phi \; ; \quad \Gamma = \Gamma_L + \Gamma_R \qquad (26)$$

where $E_F$ is Fermi energy of both leads in the unbiased junction. In several calculations a multisite bridge model was considered. In this case the electronic subsystem is represented by a linear tight-binding model and the primary phonons are modeled by assigning one local oscillator coupled to each site and to its nearest neighbor oscillators. These molecular chains are coupled at their edges to the leads electronic and phononic reservoirs. The bridge Hamiltonian is thus

$$\hat{H}_M = \sum_{i=1}^{s} \left( \varepsilon_i \hat{d}_i^+ \hat{d}_i + \omega_i \hat{a}_i^+ \hat{a}_i + M_i \hat{Q}_i^a \hat{d}_i^+ \hat{d}_i \right) + \sum_{i=1}^{s-1} \left( t_{i,i+1} \hat{d}_i^+ \hat{d}_{i+1} + U_{i,i+1} \hat{a}_i^+ \hat{a}_{i+1} + H.c. \right) \qquad (27)$$

Before presenting results of our model calculations we discuss in the following Sections two conceptual issues. First is the ambiguity in defining electron and phonon currents in the coupled system. Second is the concept of 'effective junction temperature' and its applicability in describing the non-equilibrium steady state of a current carrying junction.

## 4. Electron and phonon energy currents

Consider Eq. (17)-(19) for the phonon energy current at the molecule-lead interface. As defined this current vanishes when the interaction $U$ between primary and secondary phonons is zero. This appears contradictory to the well known fact[55-57] that energy transfer from molecular vibrations to electron-hole pair excitations is often a dominant mechanism for vibrational relaxation of molecules adsorbed at metal surfaces. Obviously a chain of coupled springs connecting with this coupling mechanism between two free electron thermal reservoirs should conduct heat via this mechanism.

The heat current defined by (17)-(19) does not contain this contribution. Indeed, in the representation that defines the Hamiltonian (2)-(3) primary phonons are not coupled directly to the outside electron reservoirs. This coupling appears only in the electronic part of the problem and should therefore be accounted as part of the electronic energy current defined by Eqs. (9), (12) and (13). It is important to realize that regarding this current as electronic or phononic is more a matter of representation than a fundamental issue of physics.

To further elucidate this point consider the model depicted in Fig. 2.

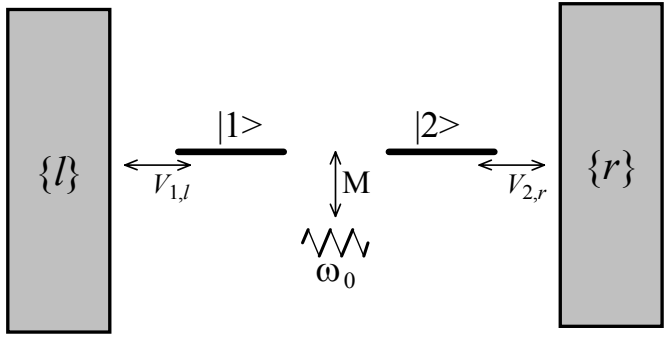

Fig. 2. A model for examining the definitions of electron and phonon energy currents where a two-electronic sites/one phonon bridge connects between two electronic reservoirs as described by the Hamiltonian of Eq.(28).

$$\hat{H} = \varepsilon_1 \hat{d}_1^\dagger \hat{d}_1 + \varepsilon_2 \hat{d}_2^\dagger \hat{d}_2 + \left(t_{12}\hat{d}_1^\dagger \hat{d}_2 + \text{h.c.}\right) + \sum_l \varepsilon_l \hat{c}_l^\dagger \hat{c}_l + \sum_r \varepsilon_r \hat{c}_r^\dagger \hat{c}_r + \omega_0 \hat{a}^\dagger \hat{a}$$
$$+ \sum_l V_{1l}\left(\hat{d}_1^\dagger \hat{c}_l + \text{h.c.}\right) + \sum_r \left(V_{2r}\hat{d}_2^\dagger \hat{c}_r + \text{h.c.}\right) + M\left(\hat{d}_1^\dagger \hat{d}_1 + \hat{d}_2^\dagger \hat{d}_2\right)\left(\hat{a}^\dagger + \hat{a}\right) \quad (28)$$

In particular we focus on the case $t_{12} = 0$ in which no electron conduction can take place. In the absence of electron-phonon coupling the electronic GFs and SEs have the block diagonal forms

$$G^r(E) = \begin{pmatrix} E - \varepsilon_1 + i\Gamma_1/2 & 0 \\ 0 & E - \varepsilon_2 + i\Gamma_2/2 \end{pmatrix}, \quad (29)$$

$$\Sigma^{>,<} = \Sigma_L^{>,<} + \Sigma_R^{>,<}, \quad (30)$$

$$\Sigma_L^< = \begin{pmatrix} i\Gamma_1 f_L(E) & 0 \\ 0 & 0 \end{pmatrix}; \quad \Sigma_R^< = \begin{pmatrix} 0 & 0 \\ 0 & i\Gamma_2 f_R(E) \end{pmatrix}, \quad (31)$$

($\Sigma_L^>$ and $\Sigma_L^>$ are similar with $if$ replaced by $-i(1-f)$)

$$G^{>,<}(E) = G^r(E)\Sigma^{>,<}(E)G^a(E), \quad (32)$$

and both the electronic current (8a) and the electronic heat flux, Eq. (15), vanish. In the presence of electron-phonon coupling the electronic current remains zero, however the heat flux is finite if the temperatures in the left and right electronic reservoirs are different. To see this we write the phonon GFs in the quasi-particle approximation whereupon they take the form of the free phonon GFs for some yet unknown distribution

$$\begin{aligned} D^<(\omega) &= -2\pi i\left[N_{ph}(\omega)\delta(\omega-\omega_0) + (1+N_{ph}(\omega))\delta(\omega+\omega_0)\right] \\ D^>(\omega) &= -2\pi i\left[N_{ph}(\omega)\delta(\omega+\omega_0) + (1+N_{ph}(\omega))\delta(\omega-\omega_0)\right] \end{aligned} \quad (33)$$

The phonon contribution to the electron SEs is given in the Born approximation by $\left[\Sigma_{ph}^{>,<}(E)\right]_{ij} = iM^2 \int (d\omega/2\pi) D^{>,<}(\omega) G_{ij}^{>,<}(E-\omega)$. This yields the block diagonal matrices

$$\Sigma_{ph}^{<}(E) = M^2\left[N_{ph}(\omega_0)G^{<}(E-\omega_0)+(1+N_{ph}(\omega_0))G^{<}(E+\omega_0)\right]$$
$$\Sigma_{ph}^{>}(E) = M^2\left[N_{ph}(\omega_0)G^{>}(E+\omega_0)+(1+N_{ph}(\omega_0))G^{>}(E-\omega_0)\right]$$
(34)

Denoting by $G_K(E)$ the 11 (K=L) or 22 (K=R) block of the full GF, Eqs. (9), (15) and (34) lead to

$$J_{Q,K}^{el} = \frac{M^2\omega_0}{\hbar}\int\frac{dE}{2\pi}G_K^{<}(E)\left[(1+N_{ph}(\omega_0))G_K^{>}(E-\omega_0)-N_{ph}(\omega_0)G_K^{>}(E+\omega_0)\right] \quad (35a)$$

in which the electronic GFs are modified by the phonon interaction, e.g. an additional term in (32) arising from (34). It is easy to check that this heat current is not zero. For example, taking in (35a) the electron-phonon interaction to the lowest ($M^2$) order leads to

$$J_{Q,K}^{el} = \frac{M^2\omega_0}{\hbar}\int\frac{dE}{2\pi}\frac{\Gamma_K^2 f_K(E)}{(E-\varepsilon_K)^2+(\Gamma_K/2)^2}$$
$$\times\left[\frac{(1+N_{ph}(\omega_0))[1-f_K(E-\omega_0)]}{(E-\omega_0-\varepsilon_K)^2+(\Gamma_K/2)^2}-\frac{N_{ph}(\omega_0)f_K(E+\omega_0)}{(E+\omega_0-\varepsilon_K)^2+(\Gamma_K/2)^2}\right]$$
(35b)

where $\varepsilon_K = \varepsilon_1(\varepsilon_2)$ for K = L (R). $N_{ph}(\omega_0)$ is now chosen so that $J_{Q,L}^{el} = -J_{Q,R}^{el}$. By the definitions of Sect. 3 this is an electron assisted heat current.

Now suppose that we first diagonalize the $\{\ell\}$-1 and 2-$\{r\}$ blocks of the electronic Hamiltonian (28). We use the same notation as before, $\{l\}$ and $\{r\}$ for the new electronic manifolds. The Hamiltonian now takes the form

$$\hat{H} = \sum_l \varepsilon_l \hat{C}_l^{\dagger}\hat{C}_l + \sum_r \varepsilon_r \hat{C}_r^{\dagger}\hat{C}_r + \omega_0 \hat{a}^{\dagger}\hat{a} + (\hat{a}+\hat{a}^{\dagger})\sum_{k,k'\in L,R} M_{kk'}\hat{C}_k^{\dagger}\hat{C}_{k'} \quad (36)$$

where $M_{kk'} \equiv v_k^* M v_{k'}$ with $|v_k|^2 = |V_{Kk}G_K^r(\varepsilon_k)|^2$. Here $K=1(2)$ for $k\in L$ (R). This representation corresponds to repartitioning the system into a new contact-bridge-contact form, obtained by a unitary transformation of the electronic basis of the problem. Within this new partitioning the bridge contains only the phonon, hence thermal transport in this picture is purely phonon-assisted. The self energy of this bridge phonon due to its coupling to the electron reservoirs is (in the Born approximation)

$$\Pi_{el,K}(\tau_1,\tau_2) = -i\sum_{k_1,k_2\in K}|M_{k_1k_2}|^2 g_{k_2}(\tau_2,\tau_1)g_{k_1}(\tau_1,\tau_2)$$
$$= -iM^2 G_K(\tau_2,\tau_1)G_K(\tau_1,\tau_2)$$
(37)

where $g_k(\tau_1,\tau_2) \equiv -i\langle T_c \hat{C}_k(\tau_1)\hat{C}_k^{\dagger}(\tau_2)\rangle$ satisfies

$$G_K(\tau_1,\tau_2) = \sum_{k \in K} |v_k|^2 g_k(\tau_1,\tau_2) \tag{38}$$

and where the coefficients $v_k$ are defined by $\hat{d}_K = \sum_{k \in K} v_k \hat{C}_k$. Eq. (37) leads to the Born approximation expression for the phonon SE projection

$$\begin{align}
\Pi_{el,K}^<(\omega) &= -iM^2 \int \frac{dE}{2\pi} G_K^<(E) G_K^>(E-\omega) \\
\Pi_{el,K}^>(\omega) &= -iM^2 \int \frac{dE}{2\pi} G_K^<(E) G_K^>(E+\omega)
\end{align} \tag{39}$$

Using these SEs in Eq. (17) leads to the (phonon-assisted thermal flux) at each contact K,

$$J_K^{ph} = -\int_0^\infty \frac{d\omega}{2\pi} \omega Tr\left[\Pi_{el,K}^<(\omega) D^>(\omega) - \Pi_{el,K}^>(\omega) D^<(\omega)\right] \tag{40}$$

At steady state these fluxes satisfy $J_L^{ph} = -J_R^{ph}$.

Obviously, Eqs. (40) and (35a) describe the same flux. Indeed, substituting (33) and (39) into (40) leads again to (35a). While (35a) was derived as a thermal flux of electronic origin, according to Eqs. (36)-(37) the flux (40) is due to coupling of the bridge phonon to electron-hole excitations in the redefined electronic reservoirs. We see that assigning heat current to electron or phonon origins can be a matter of representation.

## 5. Effective junction temperature

Theoretical discussions of junction heating often introduce the concept of 'effective junction temperature'.[16, 20, 58] This concept is obviously questionable in non-equilibrium situations, and measurable consequences of its failure were predicted.[59] On the other hand, the concept is convenient as an intuitive measure of junction heating. Usually the 'local temperature' $T_\alpha$ associated with a mode $\alpha$ of frequency $\omega_\alpha$ is introduced through the occupancy $n_\alpha$ of this mode, calculated under the given non-equilibrium conditions, by the relationship

$$n_\alpha = N_{eq}(\omega_\alpha, T_\alpha). \tag{41}$$

This definition of the effective mode temperature disregards the fact that the true vibrational distribution in the non-equilibrium system, $N_{neq}(\omega)$, can be quite different from the thermal one. The existence of the inequality

$$\sigma \equiv \int \frac{d\omega}{2\pi} |N_{neq}(\omega) - N_{eq}(\omega)| \rho_{ph}(\omega) \ll n \equiv \int \frac{d\omega}{2\pi} N_{neq}(\omega) \rho_{ph}(\omega) \tag{42}$$

may be taken as a criterion for the applicability of Eq. (41). Here $\rho_{ph}(\omega)$ is the density of bridge vibrational states which includes the effect of coupling to the phonon baths and to the tunneling electrons (and more generally, also of anharmonic phonon-phonon interactions). The condition (42) is expected to fail far from equilibrium (large source-drain voltage) and/or for strong electron-phonon coupling, $M > \Gamma$. Figure 3 shows results of a model calculation that demonstrates this breakdown. The calculation is done for a model that includes a single-level bridge (energy $\varepsilon_0$) coupled to one vibrational degree of freedom (frequency $\omega_0$), Eq. (27) with $s = 1$, using the parameters $\varepsilon_0 = 2\,\text{eV}$, $\Gamma_L = \Gamma_R = 0.02\,\text{eV}$, $\omega_0 = 0.2\,\text{eV}$, $\Omega_L = \Omega_R = 0.005\,\text{eV}$, $M = 0.2\,\text{eV}$, and $T = 100\,\text{K}$. Perhaps surprisingly (see however point (d) in the discussion of Fig. 4), the estimate based on Eq. (41) seems to be valid at high bias, however it fails quantitatively at low bias when the junction "temperature" is low. The main source of error arises from the fact that the estimate (41) is based on the equilibrium distribution for the *free* oscillator, which should indeed fail when the coupling $M$ is responsible for a substantial part of the junction energy.

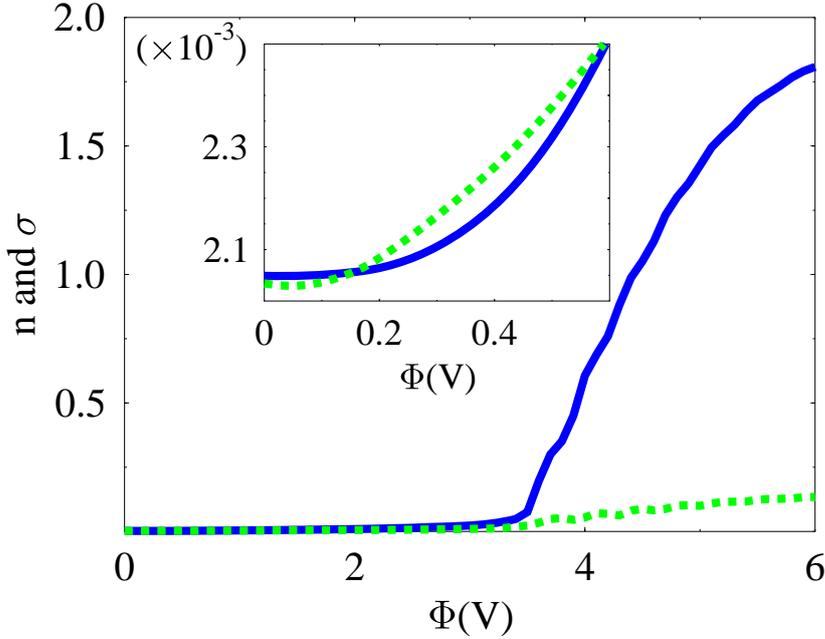

**Figure 3. Vibrational population $n$ (solid line, blue) and deviation $\sigma$ (dashed line, green) vs. the applied bias. The inset shows the low voltage region. See text for parameters. In this calculation the GFs and SEs are evaluated within the strong electron-phonon coupling scheme of Ref. [47]**

Another way to introduce an effective mode temperature can be proposed, based on the observation that experimentally temperature is measured by bringing two systems into contact and waiting till thermal equilibrium is achieved, i.e. the thermal flux between the system and the thermometer vanishes. This can be used in the following way: First, the steady state of the junction is determined using the self-consistent NEGF-based procedures of Refs. [46] (weak electron-phonon coupling) or [47] (strong coupling). Then the desired mode α is linearly coupled to an additional harmonic thermal bath (thermometer) and the temperature of the latter is adjusted so as to zero the thermal flux between the mode and the thermometer. This adjustment is made keeping the original system (bridge) and its corresponding GFs fixed. In evaluating this flux the GFs $D^{>,<}$ in Eq.(17) are due to the examined mode, while the SEs $\Pi^{>,<}$ represent the coupling of that mode to the thermometer. Note that the strength of this coupling ($\Omega_{\alpha\alpha}(\omega_a)$ from Eq. (19)) is not important since is enters linearly as multiplying factors in the flux expression. It is important however that the "thermometer" bath has a non-vanishing mode density about the examined mode. We denote this temperature as $T_{th}$.

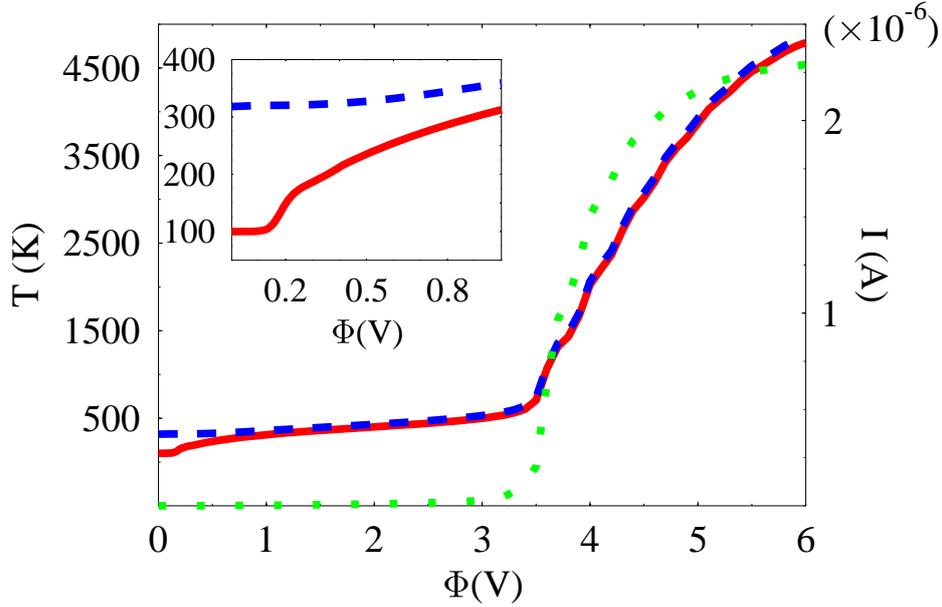

**Figure 4.** The 'local temperature' (left vertical axis) defined by the equilibrium distribution assumption, Eq. (41) (dashed line, blue) and by the measurement process explained in the text(solid line, red) plotted against the applied bias for the same model and parameters as in Fig. 3. The inset shows the low bias region. The dotted line (green) shows the junction current (right vertical axis). In this calculation the GFs and SEs are evaluated within the strong electron-phonon coupling scheme of Ref. [47]

Figure 4 compares the effective junction temperatures obtained from Eq. (41) and through the $T_{th}$ measuring approach described above. The following points are noteworthy:

(a) Junction heating, as estimated by the thermometer techniques, is characterized by two thresholds. Below the first crossover at the inelastic threshold $\Phi \sim 0.2V = \hbar\omega_0$, the junction temperature remains close to that of the leads (100K). It increases moderately above this threshold until a sharp crossover to strong heating near $\Phi \sim 4V$, where the molecular level $\varepsilon_0$ enters the conduction window in accord with Eq. (26).

(b) The two approaches to estimating $T_{eff}$ agree with each other for high bias but deviate strongly in the low bias regime where the effective temperature derived from the equilibrium distribution assumption, Eq. (41), is substantially higher than that 'measured' by the 'thermometer' bath $T_{th}$, this failure persists when $\Phi \to 0$, where the equilibrium junction must have the same temperature as that imposed on the leads, $T = 100K$

(c) As noted above, the reason for this failure Eq. (41) uses the free oscillator distribution, Eq. (20), while in fact the oscillator is not free but coupled to the electronic subsystem.

(d) Irrespective of the deviation, exemplified by Fig. 3, of the non-equilibrium distribution $N_{neq}(\omega)$ from the corresponding thermal form, estimating the effective temperature using Eq. (41) is seen to be successful (in comparison to the measuring technique) at high bias, even though the difference between $N_{eq}(\omega)$ and $N_{neq}(\omega)$ is expected to be larger. This is partially an artifact of the single resonance level/ single vibrational mode model, where results depend on properties of the phonon distribution at a relatively narrow frequency range about $\omega_0$ and not on the full non-equilibrium distribution.

## 6. Model calculations

In this Section we present several model calculations that demonstrate the application of the formalism introduced above to the issues of heat generation, heat transport and temperature rise in molecular junctions. All figures below use the SCBA scheme of Ref. [46] for the GFs and SEs calculation, except Fig. 6 where the strong electron-phonon coupling scheme of Ref. [47] was used.

(*a*) *Heat generation and temperature rise in current carrying junctions*. Consider first the heat generation rate in a current carrying molecular junction. As discussed in Section 3, this is the net rate at which energy is transferred from electrons to primary phonons on the bridge, and can be calculated from Eqs. (14) and (16). At steady state this rate reflects the non-equilibrium distributions in the electron and phonon subspaces. In the calculation of Fig. 5 we approach this issue in a slightly different way, by considering a situation in which the phonon subsystem is restricted to be in equilibrium at temperature $T_{ph}$ (that may or may not be taken equal to that of the leads) and consider the net energy transfer rate (16) under this condition as a function of $T_{ph}$ and of the bias voltage. The junction is again characterized by a single electronic level $\varepsilon_0$ and a single primary phonon $\omega_0$. The electronic GFs and SEs of this coupled electron-phonon system are calculated at the self-consistent Born approximation (SCBA)[46]. We see that the heat generation changes sign (negative sign corresponds to net energy transfer from the phonon to the electron subsystems) as a function of the imposed phonon temperature. In particular it vanishes at the equilibrium temperature (300K) of an equilibrium unbiased junction and is positive at this temperature in the biased junction indicating that the electron subsystem is in a sense "hotter".

The temperature (nearly 400K) at which heat generation vanishes in the biased junction can be identified as an effective temperature of the non-equilibrium electron subsystem.

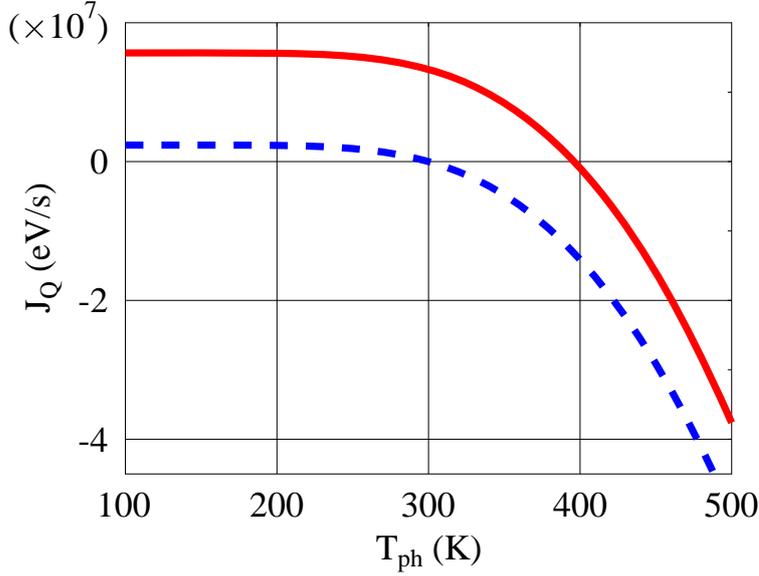

Figure 5. Heat generation in a current carrying junction characterized by one electronic level and one oscillator mode (taken to be at thermal equilibrium), plotted for a zero potential bias (dashed line; blue) and for bias $V = 0.1\,\mathrm{V}$ (solid line, red) as a function of the mode temperature. Parameters of the calculation are $\varepsilon_0 = 2\,\mathrm{eV}$, $\Gamma_L = \Gamma_R = 0.5\,\mathrm{eV}$, $E_F = 0$, $\omega_0 = 0.2\,\mathrm{eV}$ and $T_L = T_R = 300\,\mathrm{K}$.

The above calculation is similar in spirit to the measuring techniques described in Sect. 5, providing a measure of the effective electronic temperature in the non-equilibrium junction by zeroing the heat flux between it and a phonon system of known temperature, however important technical differences exist: Contrary to a model with bilinear coupling between phonon subspaces, here we deal with an approximate calculation (e.g. SCBA), so that the resulting effective temperature will depend on both the level of theory and the electron-phonon coupling strength. The computed effective temperature is expected to be meaningful only for very weak electron-phonon coupling, when the leading (second) order term in the electron phonon coupling dominates the system behavior.

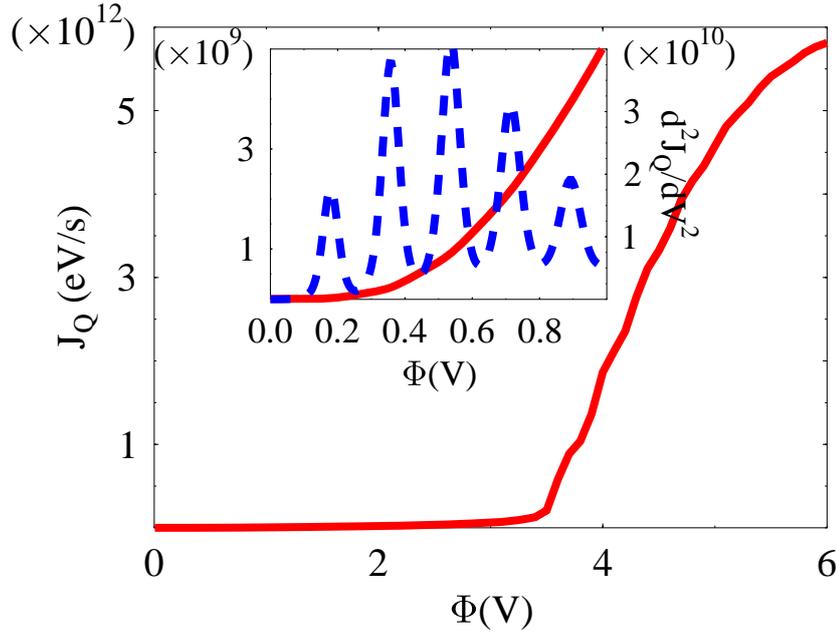

**Figure 6.** The heat generation rate, Eq. (16), in a current carrying junction characterized by one electronic level and one primary phonon coupled to a thermal bath (solid line, red) and its second derivative (dashed line, blue), plotted against the potential bias. Parameters of the calculation are $\varepsilon_0 = 2\,\text{eV}$, $\Gamma_L = \Gamma_R = 0.02\,\text{eV}$, $E_F = 0$, $\omega_0 = 0.2\,\text{eV}$, $\Omega_L = \Omega_R = 0.005\,\text{eV}$, $M = 0.2\,\text{eV}$ and $T_L = T_R = 100\,\text{K}$. The inset shows an expanded view of the low bias region. The calculation of the needed Green functions was done utilizing the strong coupling procedure of Ref. [47]

Figure 6 shows the heat generation in the current-carrying junction as a function of applied bias. The two thresholds of heat generation discussed with regard to Figure 4 are observed, the inelastic threshold at $e\Phi = \hbar\omega_0 \approx 0.2\,\text{eV}$ and the conduction threshold near $e\Phi \approx 4\,eV$ where the molecular level enters the resonance tunneling region between the left and right Fermi energies. The vibrational structure of this heat generation spectrum should be noted. It is seen as peaks in the second derivative signal about the lower $e\Phi = \hbar\omega_0$ threshold (seen in the expanded view in the inset) and as steps (that would appear as peaks in the first derivative signal) above the $e\Phi \approx 4\,eV$ threshold. The low bias behavior is characteristic of the standard inelastic tunneling spectroscopy, which is characterized by peaks in the second derivative of the current-voltage characteristic (multiple overtone peaks appear because by our choice of strong electron-phonon coupling). The high bias structure is the analog of phonon sideband peaks that often appear in the conduction-voltage plot above the conduction threshold in resonance inelastic tunneling spectroscopy. Note that the fact that here we look at the heat generation signal may affect the observed spectra. For example, higher overtones in the $d^2 I/d\Phi^2$ in IETS are rare because the probability to excite more than one phonon in a non-resonance process is small.

However higher harmonic generation corresponds to larger energy transfer to phonons, which gives larger weight to such higher harmonics peaks. This, and our choice of relatively strong electron-phonon coupling, are the probable causes for the non-monotonic shape of the signal envelope.

The balance between heat generation in the junction and heat dissipation out of the junction is expressed in terms of the steady state junction temperature. Figure 7 shows the results of calculations on a single state model (one electronic level coupled to one oscillator) where the effective temperature of this primary oscillator was determined by the measuring technique of Sect. 5. The temperature is displayed against the molecule-leads electronic coupling Γ and primary-to-secondary phonon coupling Ω. As in Fig. 5, this calculation was done using the SCBA approach to evaluate the electron and phonon GFs and SEs. As expected, stronger electronic coupling to the leads results (in the off-resonance case) in a larger junction temperature due to the higher electron flux, while stronger coupling between the bridge vibrational mode and the thermal bath of secondary phonons drives the junction towards lower temperatures closer to equilibrium with the thermal bath. Also expected is the temperature increase with the bias voltage, as seen in Fig. 4 (for slightly different junction parameters) and discussed above.

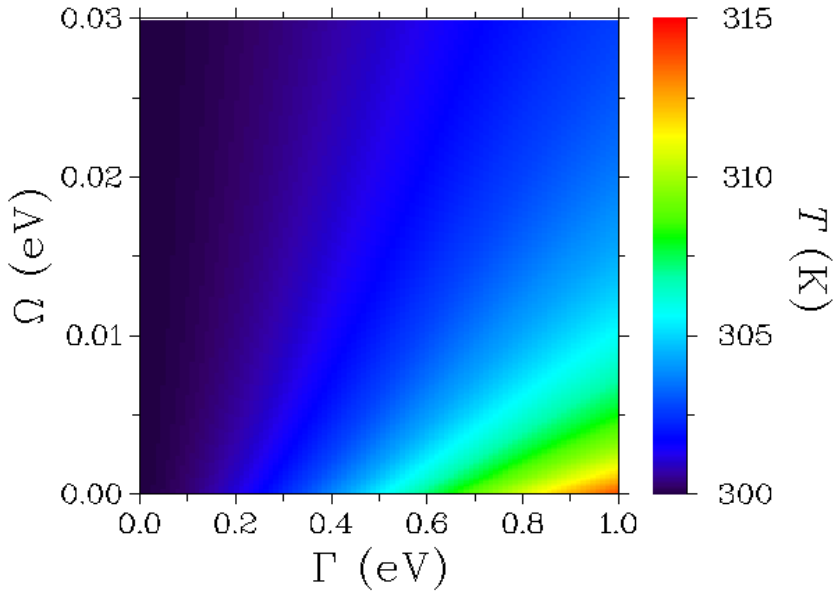

**Figure 7. Contour plot of junction temperature $T_{th}$ vs. the strength of the electronic molecule-leads coupling $\Gamma = \Gamma_L + \Gamma_R$ and the vibration-thermal bath coupling $\Omega$, in a junction characterized by symmetric electronic coupling to leads $(\Gamma_L = \Gamma_R)$ and a junction bias $\Phi = 0.1\,\text{V}$. Other junction parameters are $\varepsilon_0 = 1\,\text{eV}$, $E_F = 0$, $\omega_0 = 0.1\,\text{eV}$, $M = 0.2\,\text{eV},$ and $T_L = T_R = 300\,\text{K}.$**

(*b*) *Thermal transport.* Next consider thermal transport through the junction. We use the molecular model described above as a bridge between two metal contacts without a potential bias, i.e. $\Phi = 0$, but with a temperature bias, $T_L \neq T_R$. The ensuing process is heat conduction to which both electrons and phonons contribute. To elucidate their respective roles we first consider a junction without electron-phonon coupling, $M = 0$. In this case electron and phonon transport take place independently; the electron energy and heat currents are described by Eqs. (13) and (15) (note that in unbiased junctions and if all energies are calculated relative to the Fermi energy, these equations are identical) and the phonon current – by Eq. (24). Figures 8 show these currents, calculated for a 1-state bridge as described above. It is seen that, depending on system parameters, either phonon or electron transport can dominate the thermal current.

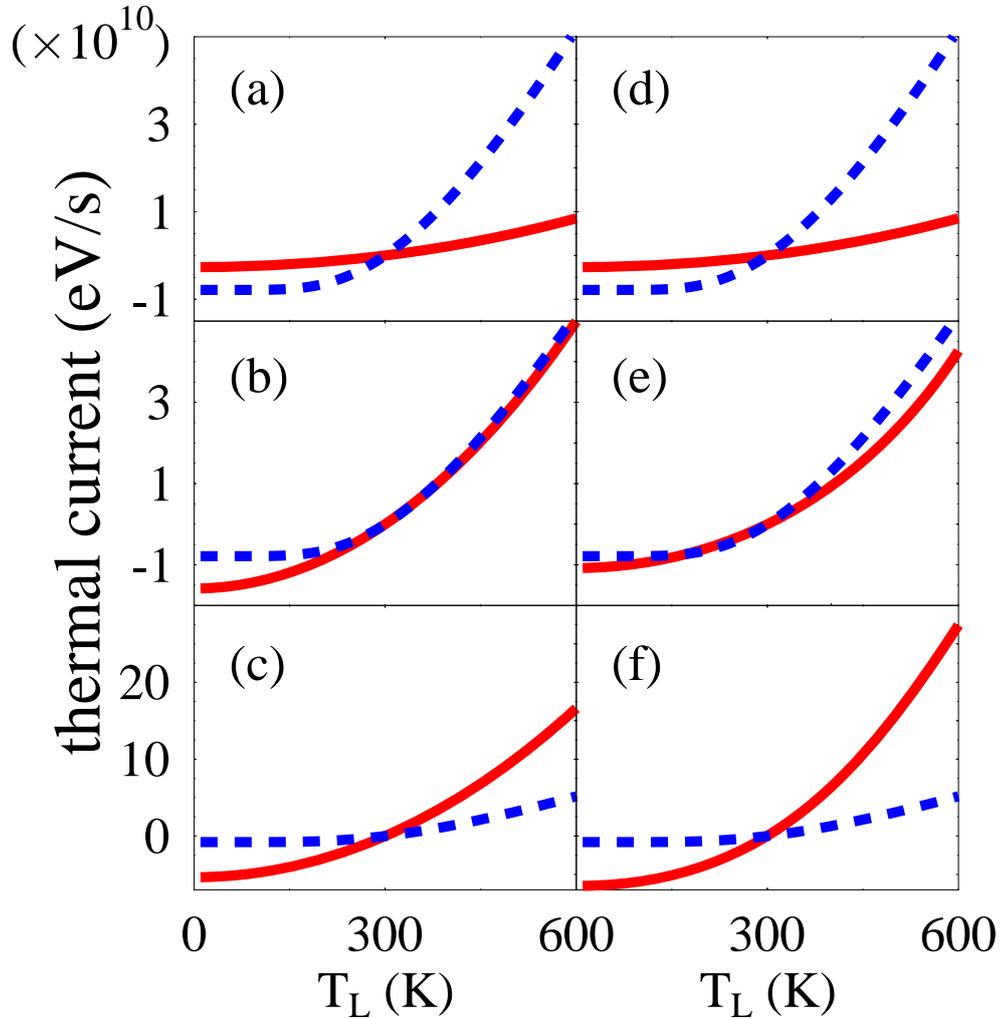

Figure 8. Thermal (energy) currents carried by electrons (solid line, red) and phonons (dashed line, blue) through a 1-state junction (see text) connecting metal leads under an imposed temperature difference between the two sides. In the calculations shown $T_R = 300\,\text{K}$ is kept fixed, while $T_L$ is varied. The junction parameters used in this calculation are $E_F = 0$, $\omega_0 = 0.1\,\text{eV}$, $\Omega_0^L = \Omega_0^R = 0.005\,\text{eV}$. In (a)-(c)

$\varepsilon_0 = 1 \text{ eV}$ and the molecule-lead electronic coupling is varied: $\Gamma_L = \Gamma_R = 0.1 \text{ eV}$ (a), $0.25 \text{ eV}$ (b), and $0.5 \text{ eV}$ (c). In (d)-(f) $\Gamma_L = \Gamma_R = 0.1 \text{ eV}$ and $\varepsilon_0$ is varied according to $\varepsilon_0 = 1 \text{ eV}$ (d); $0.5 \text{ eV}$ (e) and $0.2 \text{ eV}$ (f).

(*c*) *Thermoelectric currents*. The existence of temperature difference between unbiased metallic junctions connected by a bridge gives rise to an electric (or thermoelectric) current as well as thermal current. Figure 9 demonstrates an important characteristic of these currents. It shows both the thermal and the thermoelectric currents computed for a 1-state junction model and plotted against the temperature difference between the left and right leads. While the thermal flux is always directed from the hot to the cold contact, the direction of the thermoelectric current depends on the carrier type. Indeed, Figure 9 shows that the direction of the latter current in the case where $\varepsilon_0 - E_F = 0.5 \text{ eV}$, where the current can be characterized as electron current (solid line, red) is opposite to that obtained for $\varepsilon_0 - E_F = -0.5 \text{ eV}$ (dashed line, blue) where the dominant mechanism is hole-transport. In contrast, the electronic heat flux is the same (dotted line, black) in both cases. The dependence of the thermoelectric current on the carrier type was proposed[60] as a way to discern between electron and hole dominated transport (determined by the positions of the occupied and unoccupied molecular level relative to the leads Fermi energy) in molecular junctions.

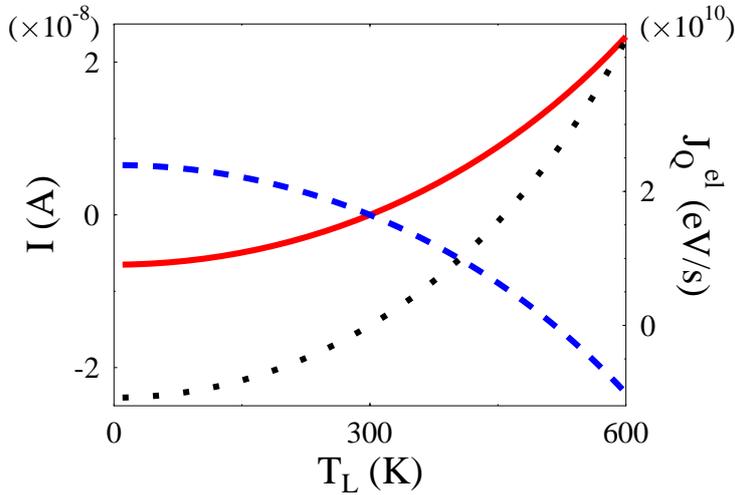

Figure 9. The thermal electron flux, $J_Q^{el} = J_{Q,L}^{el} = -J_{Q,R}^{el}$, Eq. (15), (dotted line, black; right axis) and the electric current (left axis) through a 1-level/ 1-vibration junction, plotted for an unbiased junction against the left-side temperature where the temperature on the right is kept fixed, $T_R = 300 \text{ K}$. The molecular energy level is positioned at $\varepsilon_0 - E_F = 0.5 \text{ eV}$ (solid line, red) and $-0.5 \text{ eV}$ (dashed line, blue). $\Gamma_L = \Gamma_R = 0.1 \text{ eV}$ and the other junction parameters are those of Fig. 8. In particular, electron-phonon interaction is taken zero in this calculation.

(*d*) *Effect of electron-phonon interaction*. The results shown in Figures 8 and 9 were obtained in the absence of electron-phonon interaction, i.e. $M = 0$ in the Hamiltonian (3) or (27). The significance of this coupling in thermal conduction is examined in Fig. 10. This figure shows, for the single state bridge, the total thermal flux (sum of electron and phonon heat fluxes from Eqs. (15) and (17)-(20), respectively) at the left molecule-contact interface for $M = 0.5\,\mathrm{eV}$ as well as the sum of the electron and phonon thermal fluxes, Eqs. (15) and (24) respectively, obtained for the case $M = 0$. The inset shows the difference between these results. The other two lines focus on the phonon flux (as defined in Section 3) calculated from Eq. (24), with (dotted line, green) and without (dashed line, blue) including the electron-phonon interaction in calculating the phonon Green function. Considering the electron phonon coupling that was used here is rather large, we may conclude that the effect of the electron-phonon coupling on the conduction is modest, though not negligible. It should be kept in mind however that, as discussed in Sect. 4, the definition of phonon heat flux is not unique.

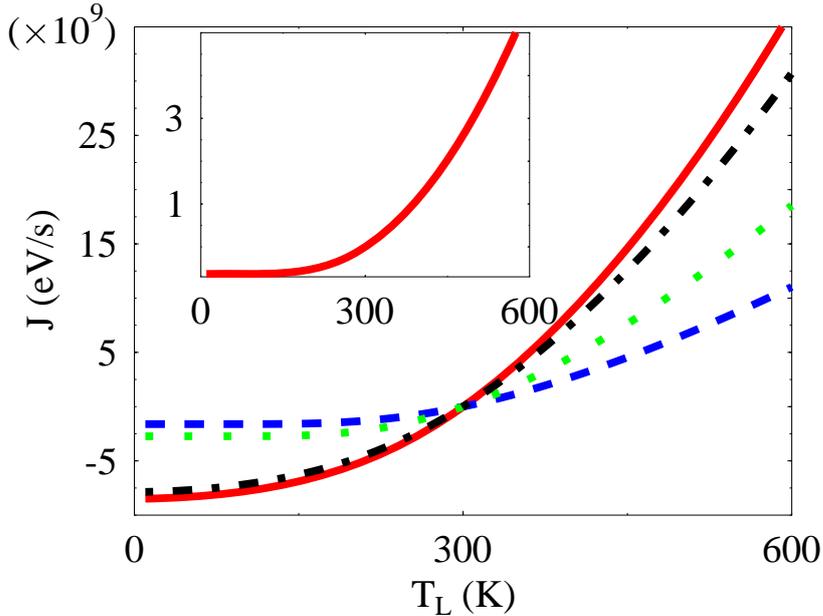

**Figure 10 Thermal flux through a single-level as a function of the temperature difference between the leads.** ($T_R = 300\,\mathrm{K}$ is kept fixed while $T_L$ is varied). The bridge parameters are $\varepsilon_0 = 1\,\mathrm{eV}$, $E_F = 0$, $\Gamma_L = 0.05\,\mathrm{eV}$, $\Gamma_R = 0.5\,\mathrm{eV}$, $\omega_0 = 0.1\,\mathrm{eV}$ and $\Omega^L = \Omega^R = 0.001\,\mathrm{eV}$. **The electron-phonon coupling is taken $M = 0.5\,\mathrm{eV}$ when present. The total flux in the presence of electron-phonon coupling is represented by the solid line (red). The sum of electron and phonon contributions to the thermal flux in the $M = 0$, case is given bt the dash-dotted (black) line. The inset shows the difference between these results. The other two lines focus on the phonon flux (as defined in Section 3) calculated from Eq. (24), with (dotted line, green) and without (dashed line, blue) including the electron-phonon interaction (M=0.5eV) in calculating the phonon Green function.**

(*e*) *Conduction by molecular chains*. Figure 11 shows the bridge length dependence of thermal flux using the molecular model of Eq. (27) and the SCBA approach. The parameters used in this calculation are $\varepsilon_0 = 1\,\text{eV}$, $\Gamma_L = \Gamma_R = 0.25\,\text{eV}$, $E_F = 0$, $\omega_0 = 0.1\,\text{eV}$, $\Omega_0^L = \Omega_0^R = 0.005\,\text{eV}$ and $T_L = 400\,\text{K}, T_R = 300\,\text{K}$. The electronic coupling between bridge sites is taken, for all $i$, to be $t_{i,i+1} = 0.1\,\text{eV}$ (dashed line, triangles, blue) and $t_{i,i+1} = 0.5\,\text{eV}$ (solid line, circles, red), and the coupling between nearest-neighbor phonons is set to $U_{i,i+1} = 0.01\,\text{eV}$. The bottom panel depicts the total thermal flux (electronic and phononic) for electron-phonon coupling strength of $M = 0.5\,\text{eV}$. The top panel shows the separate heat fluxes, electronic (in the main panel) and phononic (in the inset), for $M = 0\,\text{eV}$. For weak intersite coupling, $t = 0.1\,\text{eV}$, we see a characteristic crossover from tunneling to hopping transport when the bridge length increases. In the strong intersite coupling case, $t = 0.5\,\text{eV}$, we see a more complex behavior that is caused by the onset of resonant electron tunneling as the bridge length increases.

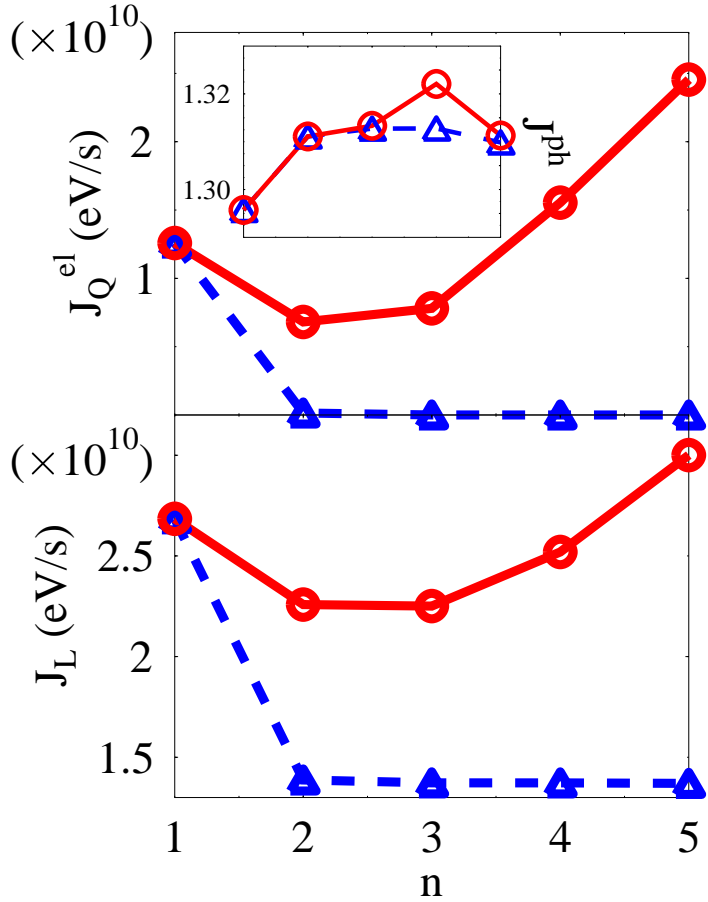

Figure 11 Thermal flux through a junction as a function of bridge length under a given temperature difference between leads. (see text for parameters). Shown are results for weak and strong intersite electronic interactions, $t = 0.1\,\text{eV}$ (dashed line, triangles, blue) and $t = 0.5\,\text{eV}$ (solid line, circles, red) (see Eq. (27)). The top panel shows the electron and phonon thermal fluxes in the absence of electron-phonon coupling, $M = 0$. The bottom panel gives the total thermal flux in the presence of strong electron-phonon coupling, $M = 0.5\,\text{eV}$.

The local temperature along the molecular chain can be examined using the measuring technique of Section 5. The result of this calculation is shown in Figure 12 for a bridge of five sites operating under potential bias of $\Phi = 0.5\,\text{V}$ (with positive right electrode; $\mu_L > \mu_R$). The parameters of this calculation are $\varepsilon_0 = 1\,\text{eV}$, $\Gamma_L = \Gamma_R = 0.1\,\text{eV}$, $E_F = 0$, $\omega_0 = 0.1\,\text{eV}$, $\Omega_0^L = \Omega_0^R = 0.025\,\text{eV}$ and $M=0.2\,\text{eV}$, with the inter-site electronic coupling set to $t = 0.5\,\text{eV}$ and the temperatures in the two leads are taken $T_L = T_R = 300\,\text{K}$. The bias potential was assumed to fall symmetrically at the two bridge-contact interfaces and stay flat along the molecular chain. With these parameters the junction temperature can become quite high. For comparison, the result obtained using the same parameters for a one-site bridge is $T \approx 319\,\text{K}$. As expected, the steady-state local temperature peaks in the interior of the bridge; sites close to contacts lose energy to the colder contacts more effectively. Interestingly, when the net electrical current goes from left to right, the steady state temperature of rightmost site is higher than that of the leftmost one. This can be rationalized using a classical picture of particles going down a slope with their kinetic energy increasing down the line. The quantum analog of this argument is that tunneling electrons that lose energy to phonons at the end of their trip through the barrier weight more in the total current because their quantum transition probability is higher than that of particles which lose energy earlier during their barrier traversal.

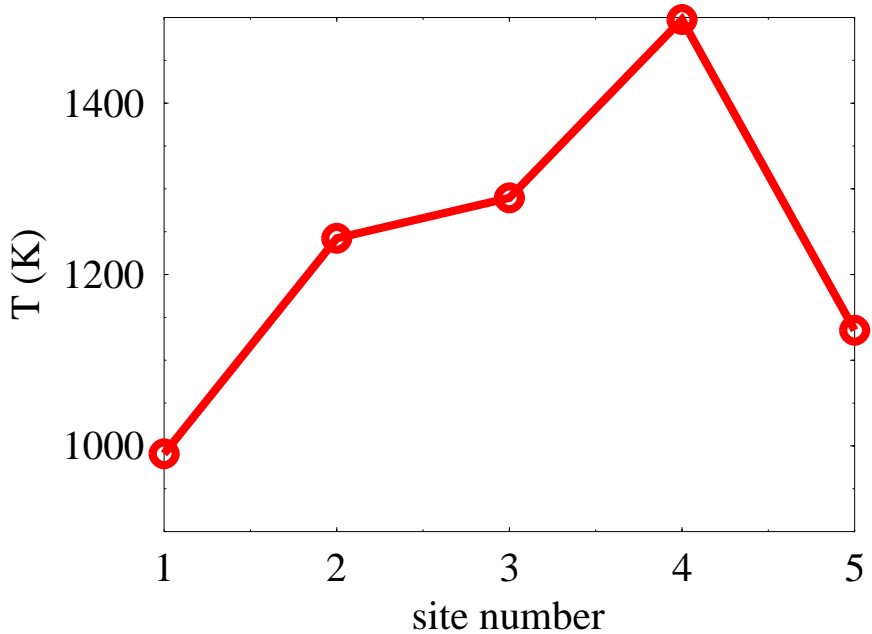

**Figure 12 The local temperature along a biased 5-sites bridge, displayed as function of the site number. See text for parameters.**

## 6. Conclusion

In this paper we have presented a general self consistent approach to thermal transport through, and heating of, a junction comprising metal electrodes connected by a molecular bridge. We employ a general non equilibrium Green function approach that makes it possible to calculate electrical and heat transport as well as heating within a unified framework that accounts self consistently for the electronic and vibrational contributions. Furthermore within the same framework we have introduced a practical definition of, and a calculation procedure for, the effective local junction temperature under non-equilibrium steady state operation. Finally, model calculations with reasonable junction parameters were used to assess the significance of junction heating, the relative contribution of electronic and vibrational degrees of freedom to the junction heat transport and the importance of accounting self consistently for the electron-phonon interaction in evaluating these thermal properties of molecular transport junctions.

## Acknowledgments

MG and MR are grateful to the DARPA MoleApps program and to the MURI program of DoD for support. The research of AN is supported by the Israel Science foundation, the US-Israel Binational Science Foundation and the Germany-Israel Foundation.

## Appendix A: Derivation of the phonon thermal flux expression

Here we derive a general NEGF expression for the phonon thermal flux through a molecule represented as system of coupled oscillators connecting two thermal phonon baths ($K = L, R$) characterized by different temperatures. This derivation essentially reproduces the derivation of Ref. [49] of the general NEGF expression for the electron flux through an electronic system connecting two thermal electron reservoirs characterized by different electrochemical potentials. It is based on the same basic assumptions: bilinear molecule-bath coupling, free carriers in the bath, and non-crossing approximation (NCA) for the molecule-bath transfer in the molecular subspace of the problem.

Consider first two bilinearly coupled classical oscillators

$$H = \sum_{i=1,2}\left[\frac{p_i^2}{2m_i} + \frac{m_i \omega_i^2 x_i^2}{2}\right] + g_{12} x_1 x_2 \qquad (43)$$

The force exerted by oscillator $i$ on oscillator $j$ is $F_{i\to j} = -g_{12} x_i$, thus the work done by $i$ on $j$ per unit time, i.e. the energy flow from $i$ to $j$ is given by

$$J_{i \to j}^{ph} = F_{i \to j} \dot{x}_j = -g_{12} x_i \frac{p_j}{m_j} \qquad (44)$$

The corresponding quantum energy flux (or thermal flux) operator is obtained from the symmetrized product. In second quantization it reads

$$\hat{J}_{i \to j}^{ph} = -\frac{U_{12} \omega_j}{2} \left[ \hat{Q}_i \hat{P}_j + \hat{P}_j \hat{Q}_i \right] \qquad (45)$$

where $U_{12} = \dfrac{\hbar g_{12}}{2\sqrt{m_1 \omega_1 \cdot m_2 \omega_2}}$ and $\hat{Q}$ and $\hat{P}$ are defined in Eqs. (4) and (5) respectively.

Now consider a molecular system $M$ represented by the Hamiltonian $\hat{H}_M$ and bilinearly coupled to thermal bath $K$. The corresponding Hamiltonian is

$$\hat{H} = \hat{H}_M + \sum_{k \in K} \omega_k \hat{a}_k^\dagger \hat{a}_k + \sum_{k \in K, m \in M} U_{km} \hat{Q}_k \hat{Q}_m \qquad (46)$$

The second term on the right hand side represents phonons in the bath and the third is the bilinear molecule-bath coupling. The thermal flux between these two subsystems is obtained from (45) by summing over the vibrational degrees of freedom in each subsystem and averaging over the corresponding distributions

$$\begin{aligned} J_{M \to K}^{ph}(t) &= \sum_{m,k} \left\langle \hat{J}_{m \to k}^{ph}(t) \right\rangle \\ &= -i \sum_{m,k} \frac{\omega_k}{2} \left[ U_{km} D_{Q_m P_k}^{<}(t,t') + U_{mk} D_{P_k Q_m}^{<}(t',t) \right]_{t=t'} \\ &= -\sum_{m,k} \omega_k \operatorname{Re}\left[ i U_{km} D_{Q_m P_k}^{<}(t,t') \right]_{t=t'} \end{aligned} \qquad (47)$$

Here and below $m$ and $k$ indicate vibrational degrees of freedom in the subspaces $M$ and $K$ respectively, and same time correlations were expressed in terms of lesser projections of (Bose operators) Green functions, $D_{AB}(\tau,\tau') = -i\left\langle T_c \hat{A}(\tau) \hat{B}^\dagger(\tau') \right\rangle$, for example $\left\langle \hat{Q}_i \hat{P}_j \right\rangle = \left[ D_{Q_i P_j}^{<}(t,t') \right]_{t=t'}$. The goal is now to express the mixed molecule-bath GFs $D_{Q_m P_k}^{<}(t,t')$ in terms of pure system GFs. This is achieved by using the identity

$$D_{Q_m P_k}(\tau,\tau') \cdot \hat{\bar{D}}_k^{-1} = \sum_{m'} \frac{U_{m'k}}{\omega_k} \frac{\partial}{\partial \tau'} D_{Q_m Q_{m'}}(\tau,\tau') \qquad (48)$$

where $\hat{\bar{D}}_k^{-1}$ is the operator (in (48) operating from the right)

$$\hat{D}_k^{-1} \equiv -\frac{1}{2}\left[ \frac{\partial^2}{\partial \tau^2} + \omega_k^2 \right] \qquad (49)$$

In integral form, and after integration by parts, Eq. (48) yields

$$D_{Q_m P_k}(\tau,\tau') = -\sum_{m'} \int_c d\tau_1\, D_{Q_m Q_{m'}}(\tau,\tau_1) \cdot \frac{U_{m'k}}{\omega_k} \frac{\partial}{\partial \tau_1} D^{(0)}_{Q_k Q_k}(\tau_1,\tau') \quad (50)$$

The superscript 0 in the contact GF $D^{(0)}_{Q_k Q_k}$ indicates that phonons in the contacts are free. These free phonon GFs can be written explicitly. The retarded, lesser and greater projections take the forms

$$\begin{aligned}
D^{(0),r}_{Q_k Q_k}(t) &= -i\theta(t)\left[e^{-i\omega_k t} - e^{i\omega_k t}\right] \\
D^{(0),<}_{Q_k Q_k}(t) &= -2\pi i\left[N_k e^{-i\omega_k t} + (1+N_k)e^{i\omega_k t}\right] \\
D^{(0),>}_{Q_k Q_k}(t) &= -2\pi i\left[(1+N_k)e^{-i\omega_k t} + N_k e^{i\omega_k t}\right]
\end{aligned} \quad (51)$$

where $N_k = N_K(\omega_k) = \left[e^{\omega_k/T_K} - 1\right]^{-1}$ is the the thermal phonon population in the bath $K$. The lesser projection of Eq. (50) onto the real time axis is

$$D^{<}_{Q_m P_k}(t,t') = -\sum_{m'} \frac{U_{m'k}}{\omega_k} \int_{-\infty}^{+\infty} dt_1$$
$$\times \left[ D^{<}_{Q_m Q_{m'}}(t,t_1) \cdot \frac{\partial}{\partial t_1} D^{(0),a}_{Q_k Q_k}(t_1,t') + D^{r}_{Q_m Q_{m'}}(t,t_1) \cdot \frac{\partial}{\partial t_1} D^{(0),<}_{Q_k Q_k}(t_1,t') \right] \quad (52)$$

using this in (47), and transforming to energy domain in the steady-state situation (where the $D(t,t') = D(t-t')$) leads to

$$J^{ph}_{M\to K} = \frac{i}{2}\int_0^\infty \frac{d\omega}{2\pi} \omega\, Tr\Big\{\Omega^K(\omega)\big[D^{<}(\omega) + D^{>}(\omega) - (2N_K(\omega)+1)(D^r(\omega) - D^a(\omega))\big]\Big\} \quad (53)$$

where $\Omega^K$ is defined in Eq. (19) and $D^{r,>,<}$ are matrices in the molecular subspace with elements $D^{r,>,<}_{mm'} \equiv D^{r,>,<}_{Q_m Q_{m'}}$). In the derivation of (53) we have used $\left[D^r_{12}(t_1,t_2)\right]^* = D^a_{21}(t_2,t_1)$, $\left[D^{>,<}_{12}(t_1,t_2)\right]^* = -D^{>,<}_{21}(t_2,t_1)$, $U_{km} = U_{mk}$, and $D^{<}_{12}(-\omega) = D^{>}_{21}(\omega)$. Eq.(17) can now be obtained from Eq.(53) using expressions (18) and defining $J^{ph}_K \equiv -J^{ph}_{M\to K}$.

## Appendix B: Another phonon bath model

In order to get a thermal flux expression analogous to (17) one has to assume that vibrational modes of the bridge are independent from each other and free (e.g. do not interact with tunneling electrons or with each other). Under this strong assumption the derivation of the phonon thermal flux expression is straightforward and goes along the same lines presented in Appendix A. The only difference is that now free phonons are those on the bridge rather than in the contacts. As a result one arrives at

$$J_K^{ph} = -\sum_\alpha \int_0^\infty \frac{d\omega}{2\pi} \omega \left[ \pi_{K,\alpha}^{ph,<}(\omega) D_\alpha^>(\omega) - \pi_{K,\alpha}^{ph,>}(\omega) D_\alpha^<(\omega) \right] \quad (54)$$

where $D_\alpha^{>,<}$ are diagonal elements of the greater and lesser phonon GFs (generally $D$ is non diagonal and disregarding non-diagonal terms corresponds to a quasi-rate-equation assumption) on the bridge and where $\pi_{K,\alpha}^{>,<}(\omega) = U_\alpha^2 S_K^{>,<}(\omega)$ with $S_K^{>,<}$ being the greater and lesser projections of the GF

$$S_K(\tau,\tau') = -i \left\langle T_c\, F\left(\{\hat{Q}_\beta^b(\tau)\}\right) F^\dagger\left(\{\hat{Q}_\beta^b(\tau')\}\right) \right\rangle; \quad \beta \in K \quad (55)$$

One can use (54) under the quasi-rate-equation assumption, to evaluate the primary phonon GF in the presence of all bridge interactions, then use its diagonal part in the expression for thermal transport in place of the free phonon GF.

**Appendix C. Derivation of Eq. (25)**

Under the assumption that no energy exchange between electron and phonon degrees of freedom on the bridge takes place, phonon current conservation implies $J_L^{ph} = -J_R^{ph}$, where each of these fluxes can be expressed by Eq. (17). This makes it possible to follow the steps of Ref. [49]. The phonon thermal flux is first written in the form

$$J^{ph} = x\, J_L^{ph} - (1-x) J_R^{ph} \quad (56)$$

where $0 < x < 1$ is an arbitrary constant. Following Ref [49] we consider the case where the functions $\Omega^L(\omega)$ and $\Omega^R(\omega)$ are proportional to each other (which always holds in the common case where their frequency dependence is disregarded). We can then choose $x = \Omega^R/\Omega$ ($\Omega = \Omega^L + \Omega^R$) and use Eq. (17) in (56) to get

$$J^{ph} = \int_0^\infty \frac{d\omega}{2\pi} \omega\, Tr\left[ \frac{\Omega^L(\omega)\Omega^R(\omega)}{\Omega(\omega)} A_{ph}(\omega) \right] (N_L(\omega) - N_R(\omega)) \quad (57)$$

where $A_{ph}(\omega) = i[D^>(\omega) - D^<(\omega)]$. Next, utilizing the non-crossing approximation, i.e.

$$D^{>,<}(\omega) = D^r(\omega)\left[\Pi_L^{>,<}(\omega) + \Pi_R^{>,<}(\omega) + \Pi_{int}^{>,<}(\omega)\right] D^a(\omega) \quad (58)$$

in Eqs. (57), and using

$$\Omega_{int}(\omega) = i\left[\Pi_{int}^>(\omega) - \Pi_{int}^<(\omega)\right]$$

leads to Eq. (25).

**References**

1    K. Schwab, E. A. Henriksen, J. M. Worlock, et al., Nature 404, 974 (2000).
2    P. Kim, L. Shi, A. Majumdar, et al., Phys. Rev. Letters 87, 215502 (2001).
3    L. Shi and A. Majumdar, Journal of Heat Transfer 124, 329 (2002).
4    D. Cahill, K. Goodson, and A. Majumdar, Journal of Heat Transfer 124, 223 (2002).
5    N. Agrait, C. Untiedt, G. Rubio-Bollinger, et al., Phys. Rev. Lett. 88, 216803 (2002).
6    D. Cahill, W. K. Ford, K. E. Goodson, et al., J. Appl. Phys. 93, 793 (2003).
7    M. J. Montgomery, T. N. Todorov, and A. P. Sutton, J Phys.: Cond. Matter 14, 5377 (2002).
8    B. N. J. Persson and P. Avouris, Surface Science 390, 45 (1997).
9    A. P. van Gelder, A. G. M. Jansen, and P. Wyder, Phys. Rev. B 22, 1515 (1980).
10   A. Cummings, M. Osman, D. Srivastava, et al., Phys. Rev. B 70, 115405 (2004).
11   I. Paul and G. Kotliar, Phys. Rev. B 67, 115131 (2003).
12   D. Segal, A. Nitzan, and P. Hanggi, J. Chem. Phys. 119, 6840 (2003).
13   D. Segal and A. Nitzan, Phys. Rev. Lettters 94, 034301 (2005).
14   D. Segal and A. Nitzan, J. Chem Phys. 122, 194704 (2005).
15   K. R. Patton and M. R. Geller, Phys. Rev. B 64, 155320 (2001).
16   Y. C. Chen, M. Zwolak, and M. Di Ventra, Nano Lett. 3, 1691 (2003).
17   S. Datta, J. Phys.: Cond. Matter 2, 8023 (1990).
18   R. K. Lake and S. Datta, Phys. Rev. B 45, 6670 (1992).
19   R. K. Lake and S. Datta, Phys. Rev. B 46, 4757 (1992).
20   D. Segal and A. Nitzan, J. Chem. Phys. 117, 3915 (2002).
21   Z. Rieder, J. L. Lebowitz, and E. Lieb, J. Chem. Phys. 8, 1073 (1967).
22   A. Casher and J. L. Lebowitz, J. Math. Phys. 12, 1701 (1971).
23   A. J. O'Connor and J. L. Lebowitz, J. Math. Phys. 15, 692 (1974).
24   F. Mokross and H. Buttner, J.Phys. C 16, 4539 (1983).
25   U. Zürcher and P. Talkner, Phys. Rev. A 42, 3278 (1990).
26   S. Lepri, R. Livi, and A. Politi, Phys. Rev. Lett. 78, 1896 (1997).
27   B. Hu, B. Li, and H. Zhao, Phys. Rev. E 57, 2992 (1998).
28   L. G. C. Rego and G. Kirczenow, Phys. Rev. Letters 81, 232 (1998).
29   D. M. Leitner and P. G. Wolynes, Phys. Rev. E 61, 2902 (2000).
30   M. Terraneo, M. Peyrard, and G. Casati, Phys. Rev. Lett. 88, 094302 (2002).
31   B. Li, L. Wang, and G. Casati, Phys. Rev. Letters 93, 184301 (2004).
32   M. Kindermann and S. Pilgram, Phys. Rev. B 69, 155334 (2004).
33   S. Y. Cho and R. H. McKenzie, Phys. Rev. B 71, 045317 (2005).

[34] B. Li, J. Lan, and L. Wang, Phys. Rev. Lett. 95, 104302 (2005).

[35] K. Saito, J. Phys. Soc. Japan 75 034603 (2006).

[36] L. Y. Gorelik, A. Isacsson, M. V. Voinova, et al., Phys. Rev. Lett. 80, 4526–4529 (1998).

[37] A. Y. Smirnov, L. G. Mourokh, and N. J. M. Horing, Phys. Rev. B 67, 115312 (2003).

[38] K. D. McCarthy, N. Prokof'ev, and M. T. Tuominen, Phys. Rev. B 67, 245415 (2003).

[39] Y. M. Blanter, O. Usmani, and Y. V. Nazarov, Phys. Rev. Letters 93, 136802 (2004).

[40] N. M. Chtchelkatchev, W. Belzig, and C. Bruder, Phys. Rev. B 70, 193305 (2004).

[41] J. Koch and F. v. Oppen, Phys. Rev. Letters 94, 206804 (2005).

[42] J. Koch and F. von Oppen, Phys. Rev. B 72, 113308 (2005).

[43] J. Koch, M. Semmelhack, F. v. Oppen, et al., Phys. Rev. B 73, 155306 (2006).

[44] S. Datta, *Electric transport in Mesoscopic Systems* (Cambridge University Press, Cambridge, 1995).

[45] A. Mitra, I. Aleiner, and A. Millis, Phys. Rev. B 69, 245302 (2004).

[46] M. Galperin, M. A. Ratner, and A. Nitzan, J. Chem. Phys., 11965 (2004).

[47] M. Galperin, A. Nitzan, and M. A. Ratner, Phys. Rev. B 73, 045314 (2006).

[48] G. D. Mahan, *Many-particle physics* (Plenum press, New York, 2000).

[49] Y. Meir and N. S. Wingreen, Phys. Rev. Lett. 68, 2512 (1992).

[50] H. Haug and A.-P. Jauho, *Quantum Kinetics in Transport and Optics of Semiconductors* (Springer, Berlin, 1996).

[51] J.-S. Wang, J. Wang, and N. Zeng, Phys. Rev. B 74, 033408 (2006).

[52] N. Mingo, Phys. Rev. B 74, 125402 (2006).

[53] A. Nitzan, *Chemical Dynamics in Condensed Phases* (Oxford University Press, Oxford, 2006), Chapter 13.

[54] A. Ozpineci and S. Ciraci, Phys. Rev. B 63, 125415/1 (2001).

[55] M. Galperin, M. A. Ratner, and A. Nitzan, Nano Letters 4, 1605 (2004).

[56] A. Baratoff and B. N. J. Persson, J. Vac. Sci. & Tech. A 6, 331 (1988).

[57] V. Krishna and J. C. Tully, J. Chem. Phys. 125, 054706 (2006).

[58] D. Segal and A. Nitzan, Phys. Rev. E 73, 026109 (2006).

[59] M. Hartmann and G. Mahler, Europhys. Lett. 70, 579 (2005).

[60] M. Paulsson and S. Datta, Phys. Rev. B 67, 241403 (2003).